\documentclass[12pt,twoside]{article}

\usepackage{amsmath,amsthm,amssymb,euscript,amscd,graphicx,epsfig,color, amsfonts, fleqn }

\definecolor{mybackgroundcolor}{rgb}{1,0,1}
\definecolor{plum}{rgb}{.5,0,.5}
\definecolor{dkred}{rgb}{.5,0,0}
\definecolor{ddkred}{rgb}{.35,0,0}
\definecolor{dkblue}{rgb}{.1,0,.6}
\definecolor{ddkblue}{rgb}{0,0,.25}
\definecolor{dkgreen}{rgb}{0,.5,0}
\definecolor{ddkgreen}{rgb}{0,.35,0}
\definecolor{dkgr2}{rgb}{0,.57,0}
\definecolor{dkgr3}{rgb}{0,.64,0}
\definecolor{dkgr4}{rgb}{0,.71,0}

\pagecolor{white}

\newcommand{\be}{\begin{equation}}
\newcommand{\ee}{\end{equation}}
\newcommand{\bd}{\begin{displaymath}}
\newcommand{\ed}{\end{displaymath}}
\newcommand{\bea}{\begin{eqnarray}}
\newcommand{\eea}{\end{eqnarray}}

\newcommand{\Hi} {{\cal H}^\infty}

\begin{document}
\title{Controller Tuning for Active Queue Management Using a Parameter Space Method\thanks{This work was supported
 in part by the National Science Foundation under grant ANI-0073725}}
\author{Murat Sa\u{g}lam$^{\dag}$, Sami Ezercan\thanks{Department of Electrical and Electronics Engineering,
Bilkent University, Bilkent, Ankara, TR $68000$, Turkey;
\textit{email:} saglamt@ug.bilkent.edu.tr, sami@ug.bilkent.edu.tr,
hitay@bilkent.edu.tr}, Suat G\"um\"u\c{s}soy$^{\ddag}$ and Hitay
\"Ozbay$^{\dag}$~\thanks{Department of Electrical Engineering, The
Ohio State University, Columbus, OH $43210$, USA \textit{email:}
gumussos@ee.eng.ohio-state.edu, ozbay@ee.eng.ohio-state.edu}}
\date{}
\maketitle

\begin{abstract}
In recent years, different mathematical models have been
proposed for widely used internet control mechanisms.
Simple low order controllers (such as PID, and Smith predictor based
linear controllers that are easy to implement) are desired for network
traffic management. In order to design such simple controllers for
Active Queue Management (AQM), delay based linear models have been
considered. In this paper we discuss tuning of the PID controllers by using
a parameter space method, which computes stability regions of a class of
quasi-polynomials in terms of free controller parameters.
\end{abstract}

\section{Introduction}
\setcounter{equation}{0}

Several active queue management (AQM) schemes supporting
transmission control protocol (TCP) have been proposed in literature,
\cite{MGT00,HMTG02,K01,M03,RKH03}. Simple low order
controllers are desired for implementation purposes. Such
controllers are considered in
\cite{HMTG02,M03,QO03,FRL03,RRQ03,PZ03} for AQM.
In  particular, in \cite{HMTG02}, the TCP congestion avoidance mode
is modeled by delay differential equations with a nonlinearity, and a
PI controller is proposed for control mechanism.
Although the controller design guarantee
some robustness for parametric uncertainties, the high frequency
dynamics are considered as parasitics. We will take into account
the plant structure and design the PID controller without
simplification (except for the linearization of TCP).

In this paper, tuning of the PID controllers are discussed.
By using a parameter space method \cite{HA03}, we
compute stability regions of a class of quasi-polynomials in terms
of free controller parameters \cite{OS03}. We find the optimal PID
controller parameters by minimization of mixed sensitivity
function in stability region.

In section \ref{sec:model}, the
mathematical model of the AQM schemes and the linearized plant
to be controlled are summarized from \cite{HMTG02,QO03}.
The region of PID parameters achieving a stable close loop is found
and optimal parameter search resulting robust stability and good
performance is proposed in section \ref{sec:tuning}.
Section \ref{sec:simulations} gives the simulation results and
discusses robustness and performance of controller.
The paper ends with concluding remarks.

\section{Mathematical Model of AQM Scheme Supporting TCP Flows} \label{sec:model}

We assume that the network configuration is the same as discussed in
\cite{HMTG02,QO03}, i.e., network has a single router receiving
$N$ TCP flows. We consider  the congestion avoidance mode only.
The TCP slow start and time out mechanisms are ignored.

When $N$ TCP flows are interacting with a single router,
additive-increase multiplicative-decrease behavior of the congestion avoidance
mode has been modeled in \cite{MGT00} by the differential equation
\be
dW(t)=\frac{dt}{R(t)}-\frac{W(t)}{2}dN(t)
\ee where $R(t)=\frac{q(t)}{C}+T_p$ and other variables are defined as:
\bea
\nonumber q(t)&\doteq&\text{queue length at router,} \\
\nonumber W(t)&\doteq&\text{congestion window size,} \\
\nonumber R(t)&\doteq&\text{round trip time delay,} \\
\nonumber dN(t)&\doteq&\text{number of marks the flow suffers,} \\
\nonumber T_p&\doteq&\text{propagation delay,} \\
\nonumber C&\doteq&\text{router's transmission capacity.}
\eea
For $N$ homogeneous TCP sources and one router, nonlinear model of AQM implementation is given in \cite{HMTG02} as
\bea
\dot{W}(t)&=&\frac{1}{R(t)}-\frac{W(t)}{2}\frac{W(t-R(t))}{R(t-R(t))}p(t-R(t)) \label{eq:qdW} \\
\dot{q}(t)&=&\left[\frac{N(t)}{R(t)}W(t)-C\right]^+ \label{eq:qdq}
\eea
where $p(t)$ is the probability of packet mark used by AQM mechanism at the router and
\bd
[x]^+\doteq\left\{\begin{array}{cc}
  x & x\geq0 \\
  0 & x<0
\end{array}\right. .
\ed

The equations (\ref{eq:qdW}) and (\ref{eq:qdq}) can be linearized
about the operating point ($R_0, W_0, p_0$).
The operating point is defined by the following equilibrium
conditions (see \cite{HMTG02})
\bea
R_0&=&\frac{q_0}{C}+T_p, \\
W_0&=&\frac{R_0C}{N}, \\
p_0&=&\frac{2}{W_0^2}.
\eea
Note that the implicit nonlinearity ($t-R(t)$) is approximated by ($t-R_0$)
in the linearization of (\ref{eq:qdW}) and (\ref{eq:qdq}).

\setlength{\unitlength}{1mm}
\begin{center}
\begin{picture}(125,30)(0,30) \label{fig:overall}
\thicklines
\put(30,40){\framebox(25,12)[c]{$C(s)$}}
\put(80,40){\framebox(25,12)[c]{$P(s)$}} \put(67,47){\circle{5}}
\put(55,47){\vector(1,0){9}} \put(70,47){\vector(1,0){10}}
\put(67,56){\vector(0,-1){6}} \put(61,43){$-$} \put(63,50){$+$}
\put(69,54){$u$} \put(105,47){\vector(1,0){15}} \put(122,45){$y$}
\put(112,47){\line(0,-1){15}} \put(112,32){\line(-1,0){93}}
\put(19,47){\circle{5}} \put(21.5,47){\vector(1,0){7.9}}
\put(19,32){\vector(0,1){12.5}} \put(6.5,47){\vector(1,0){10}}
\put(3,45){$r$} \put(23,50){$e$} \put(15,41){$-$} \put(14,49){$+$}
\end{picture}\\
Figure \ref{fig:overall}. The feedback system
\end{center}

The linearized model of the plant is given in \cite{HMTG02} as
follows: \be \label{eq:plantP}
P(s)=\frac{N_p(s)}{D_p(s)}=\frac{Ke^{-R_0s}}{W_0R_0^2s^2+\left(W_0+1\right)R_0s+2+R_0se^{-R_0s}}.
\ee
where $K=\frac{NW_0^3}{2}$.
Given the plant, $P$,  a PI controller, $C$, is proposed in
\cite{HMTG02} and an $\Hi$ controller is suggested in \cite{QO03} for
the feedback loop in Figure \ref{fig:overall}. We will consider in
this paper a PID controller which has a better performance/robustness
than a PI controller and simple structure for implementation compared
to $\Hi$ controller. A method for tuning of the PID parameters will be
given in the next sections. Also, the high frequency dynamics of the
system (the delay term in the denominator of $P$) is considered as
parasitic dynamics in PI controller design \cite{HMTG02}.
We will include these effects in the design of PID controllers.

\section{Tuning of the PID Parameters} \label{sec:tuning}
In this section, we will define the controller parameter space such
that closed loop system is stable. We will use the  method proposed in
\cite{HA03}. This approach separates the parameter space of PID
controller into \textit{stable} and \textit{unstable} region. The
stability of a region is checked by the direct method in \cite{OS03}.
Once the ``stable region'' determined, the optimal parameters are
obtained by a numerical search algorithm: the criterion used here is the
minimization of the mixed sensitivity function in the region of
admissible parameter values.

\subsection{The Feasible PID Parameter Space}
For the plant, $P$, of the closed loop system Figure~\ref{fig:overall} is
to be stabilized by the PID controller,
\be \label{eq:contPID}
C(s)=K_P+K_Ds+\frac{K_I}{s}.
\ee
The triplet $(K_P,K_I,K_D)$
stabilizes the overall system if and only if all the
roots of the characteristic equation,
\bea \label{eq:CE}
CE(s)&=&s D_p(s)+(K_I+K_Ps+K_Ds^2) N_P(s) \\
\nonumber &=&(W_0R_0s^3+(W_0+1)R_0s^2+2s)
+(KK_I+KK_Ps+(KK_D+R_0)s^2)e^{-R_0s},
\eea
lie in the open left half plane.
The algorithm in \cite{HA03} offers a parameter space
approach to certain class of quasi-polynomials in the form of
\be \label{eq:GCE}
G(s)=(r_0+r_1s+r_2s^2)A(s)+B(s)e^{sL}, \quad \text{$L>0$}
\ee
where $A$ and $B$ are polynomials with degrees $m$ and $n$ respectively
satisfying $n\geq m+2$. It computes the ``stable region'' for the
triplet $(r_0,r_1,r_2)$.
We form the quasi-polynomial as the characteristic equation
of our time delayed system in (\ref{eq:CE}) as
\bea \label{eq:paramtrans}
r_0&=&KK_I, \\
\nonumber r_1&=&KK_P, \\
\nonumber r_2&=&KK_D+R_0, \\
\nonumber A(s)&=&1, \\
\nonumber B(s)&=&W_0R_0s^3+(W_0+1)R_0s^2+2s, \\
\nonumber L&=&R_0.
\eea

As explained in \cite{HA03}, a stable quasi-polynomial will be
unstable only when a left half plane root transients to right half
plane. Since $K_P$, $K_D$, $K_I$ and $h$ change continuously,
characteristic equation also changes continuously. Thus, for some
$(K_P,K_D,K_I)$ triplet, some roots of (\ref{eq:CE}) lie on
imaginary axis. From these $(K_P,K_D,K_I)$, we can form the
stability boundaries in the parameter space. In \cite{HA03}, these
crossings are classified into 3 cases and for our problem the
boundaries can be found as:
\begin{enumerate}
\item \textit{Real Root Boundary} (RRB),
a root crosses imaginary axis at origin,
i.e., $G(0)=0$, the boundary is $r_0=0$ line, equivalently, $K_I=0$.
\item \textit{Infinite Root Boundary} (IRB)
when a root crosses the imaginary axis at infinity,
since $m=0$ and $n=3$, the quasi-polynomial is
retarded type $(n>m+2)$ and no infinite root boundary exists.
\item \textit{Complex Root Boundary} (CRB)
when a pair of complex conjugate roots crosses
the imaginary axis, i.e., $G(j\omega)=0$, then
we can separate real and imaginary parts as,\\
\bd
\hspace*{-1cm}\left[\begin{array}{cc}
  1 & -\omega^2 \\
  0 & 0 \\
\end{array}\right]\left[\begin{array}{c}
  r_0 \\
  r_2 \\
\end{array}\right]+\left[\begin{array}{c}
  \omega((W_0R_0^2\omega^2-2)\sin{R_0\omega-(W_0+1)R_0\omega\cos{R_0\omega})} \\
  -\omega(W_0R_0^2\omega^2-2)\cos{R_0\omega-(W_0+1)R_0\omega\sin{R_0\omega}+r_1)} \\
\end{array}\right]=\left[\begin{array}{c}
  0 \\
  0 \\
\end{array} \right]
\ed
If we fix $r_1=r_1^{*}$, the solution of the above system of equations
exists only for real zeros of $\omega_{gi}$ of
\be \label{eq:gw}
g(\omega)=\omega r_1^{*} +(W_0R_0^2\omega^2-2)\omega\sin{R_0\omega}-(W_0+1)R_0\omega^2\cos{R_0\omega}.
\ee
Each positive zero corresponds a straight line as CRB
in $(r_2,r_0)$ plane with equation,
\be
r_0=\omega_{gi}^2 r_2-\omega_{g}((W_0R_0^2\omega_{g}^2-2)\sin{R_0\omega_{g}-(W_0+1)R_0\omega_{g}\cos{R_0\omega_{g}})}.
\ee
\end{enumerate}
Stability boundaries (RRB, IRB, CRB) can be found as explained above.
For each $r_1$, we can find a region in $(r_2,r_0)$ plane,
and for various $r_1$ values, we can form a three dimensional
region, $\Pi_r$. Any triplet in this region as PID parameters
(i.e., $(r_0,r_1,r_2)\in\Pi_r$) ensures stable closed loop system.
Note that actual PID parameters are not the triplet $(r_0,r_1,r_2)$,
but $(K_P,K_I,K_D)$. After the region, $\Pi_r$, obtained,
we can transform to another region, $\Pi_K$, for the triplet $(K_P,K_I,K_D)$
by linear transformation in (\ref{eq:paramtrans}).
Also, in order to check the inside of the region, $\Pi_K$,
(whether it is stable or not) we used the method in \cite{OS03},
which is not discussed here.

\subsection{Computation of the Optimal PID Parameters}
We aim to find the optimal PID parameters such that we have robust stability
and good tracking for the set point changes. For this purpose we will
use the weighted $\Hi$ norm of mixed sensitivity function, as our
performance metric. Since a PID controller has three parameters,
$(K_P,K_I,K_D)$, we can search the optimal triplet,
$(K_{P,opt},K_{I,opt},K_{D,opt})$ in $\Pi_K$
such that the mixed sensitivity function,
\be
\Psi(K_P,K_I,K_D)=\sup_{\omega\in[0,\infty)}
\left\{|W_1(j\omega)S(j\omega)|^2+|W_2(j\omega)T(j\omega)|^2\right\},
\ee
attains its minimum value.
The weight functions $W_1$ and $W_2$ are finite dimensional terms
as  design parameters for robustness and performance.
The sensitivity, $S$, and complementary sensitivity, $T$, functions
are defined as usual:
\bea
\nonumber S(s)&=&(1+P(s)C(s))^{-1} \\
\nonumber T(s)&=&P(s)C(s)(1+P(s)C(s))^{-1}
\eea
where $P$ and $C$ are given in (\ref{eq:plantP}) and (\ref{eq:contPID}),
respectively. Formally, we can define the numerical search
problem as:
\textit{Find the triplet, $(K_{P,opt},K_{I,opt},K_{D,opt})$ such that
\be
\Psi(K_{P,opt},K_{I,opt},K_{D,opt})\leq\Psi(K_P,K_I,K_D), \quad \forall\;(K_P,K_I,K_D)\in\Pi_K,
\ee
the inequality is satisfied.} We will simply use a brute force method by
taking a grid set in the feasible parameter space.

\section{Simulation Results} \label{sec:simulations}
We have simulated the nonlinear model defined by equations
(\ref{eq:qdW}) and (\ref{eq:qdq}) for the dynamics of $N$ TCP
flows loading a router by using \verb"simulink" and \verb"MATLAB".
The numerical values for the simulations are taken to be the
same as in \cite{QO03} for comparison purposes:
\begin{itemize}
    \item Nominal values known to the controller: $N_n=50$ TCP
    sessions, $C_n=300$ packets/sec, $T_p=0.2$ sec.
    Then, by simple calculation $R_{0n}=0.533$ sec and $W_{0n}=3.2$
    packets. Desired queue length is $q_0=100$ packets.
    \item Real values of the plant: $N=40$ TCP sessions, $C=250$
    packets/sec, $T_p=0.3$ sec, which means that we actually have
    $R_0=0.7$~sec and $W_0=4.375$ packets.
\end{itemize}
The above data will be used to check the performance of the overall
feedback system. In order to analyze the robustness of closed loop
system with respect to variations in the network parameters, the
following scenario is considered: outgoing link capacity, $C$, is
a normally distributed random signal with mean $250$ packets/sec
and variance $50$ added to a pulse of period $60$sec, amplitude
$60$ packets/sec. The number of TCP flows $N$ is a normally
distributed random signal with mean $45$ and variance $30$ added
to a pulse of period $20$ sec and amplitude $10$. The propagation
delay $T_p$ is a normally distributed random signal with mean
$0.8$ sec and variance $0.05$ sec added to a pulse of period $20$
sec and amplitude $0.2$ sec. The controllers have the following
values known to them: $C=300$ packets/sec, $N=50$, $T_p=0.7$sec
and desired queue length is $q_0=100$ packets.

\subsection{Tuning PID Parameters}
For the given network parameters, we can write the characteristic
equation from (\ref{eq:GCE}),
\bea
G(s)&=&(r_0+r_1s+r_2s^2)A(s)+B(s)e^{sL} \\
\nonumber &=&(r_0+r_1s+r_2s^2)+(1.706s^3+2.239s^2+2s)e^{0.533s} \\
\nonumber &=&(819.2K_I+819.2K_Ps+(819.2K_D+0.533)s^2)+(1.706s^3+2.239s^2+2s)e^{0.533s}
\eea
We will work with $(r_0,r_1,r_2)$ triplet and calculate the PID parameters, $(K_P,K_D,K_I)$, at the end.
It is clear that for this plant $m=0$ and $n=3$. Also note that
$B(s)$ does not contain any constant term. Therefore, we do not
encounter any infinite root boundary (IRB) and have always a real
root boundary (RRB) which is $r_0=0$.

\begin{figure}[ht]
\begin{center}
\begin{minipage}[b]{0.5\textwidth}
\centering
\includegraphics[width=3.1in,height=2.5in]{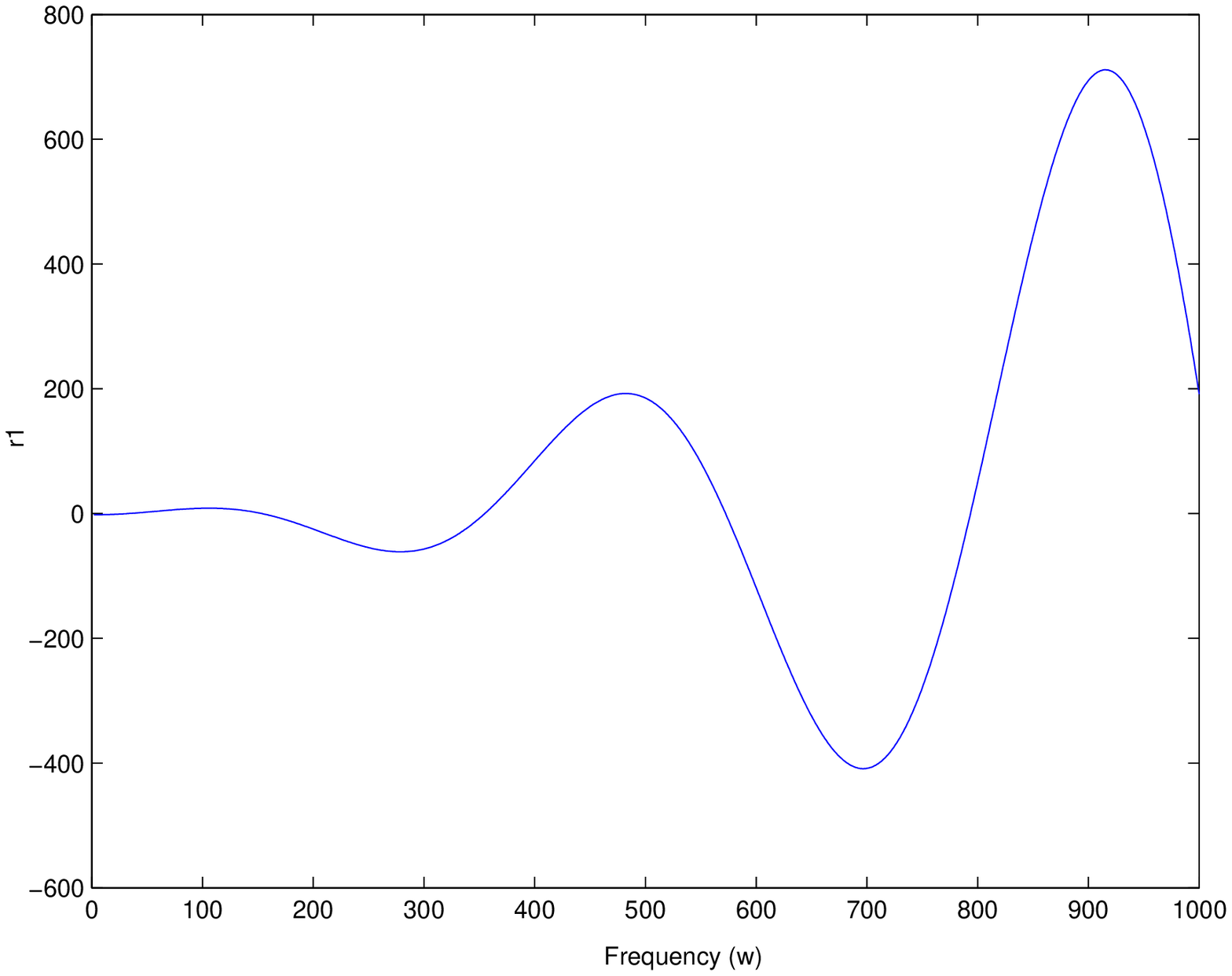}
\caption{Plot of $g(\omega)$}
\label{fig:r1_interval}
\end{minipage}%
\begin{minipage}[b]{0.5\textwidth}
\centering
\includegraphics[width=3in,height=2.45in]{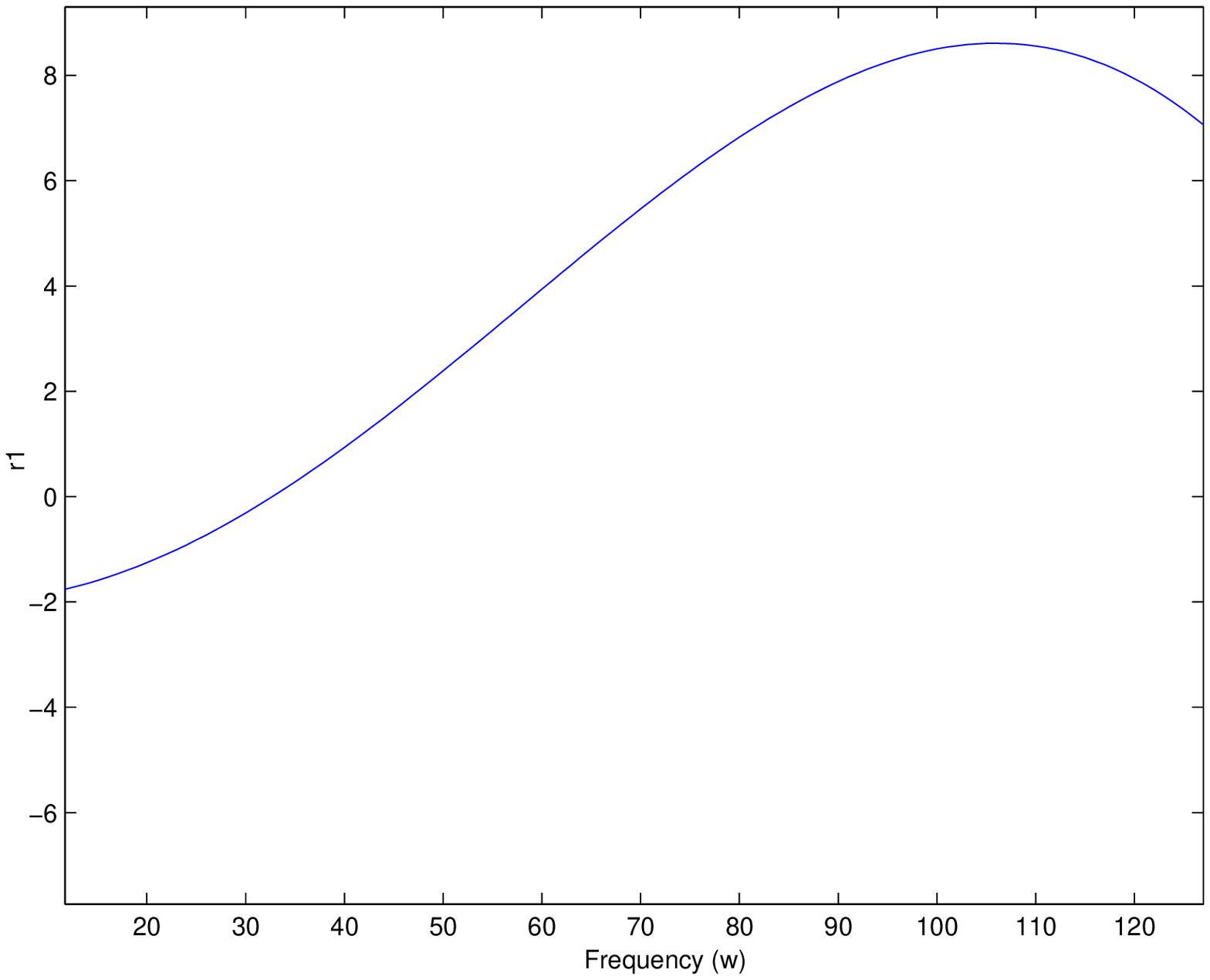}
\caption{Plot of $g(\omega)$ zoomed}
\label{fig:r1_interval_zoom}
\end{minipage}
\begin{minipage}[b]{0.5\textwidth}
\centering
\includegraphics[width=3in,height=2.35in]{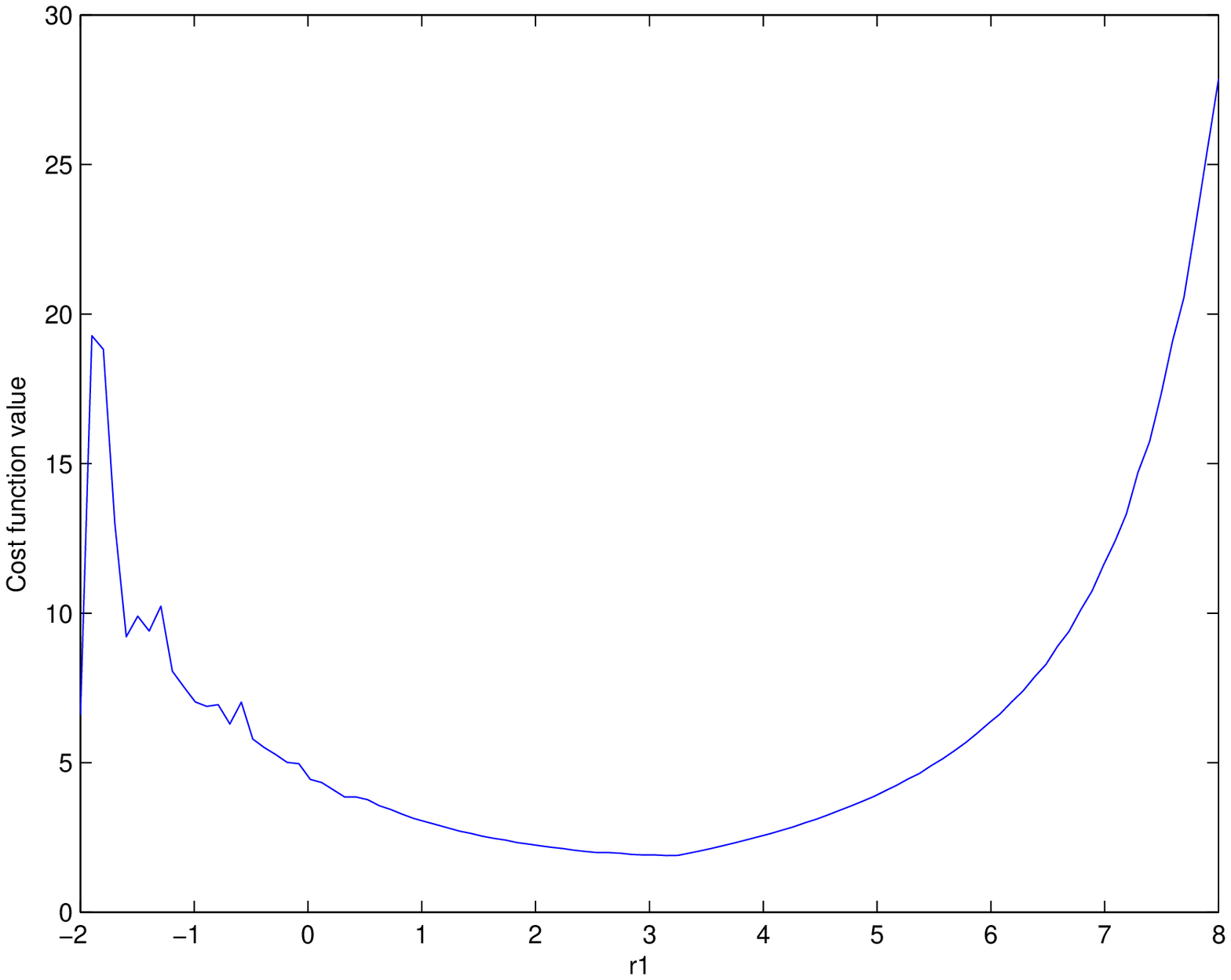}
\caption{Plot of $\Psi$}
\label{fig:r1_vs_cost}
\end{minipage}%
\begin{minipage}[b]{0.5\textwidth}
\centering
\includegraphics[width=3in,height=2.45in]{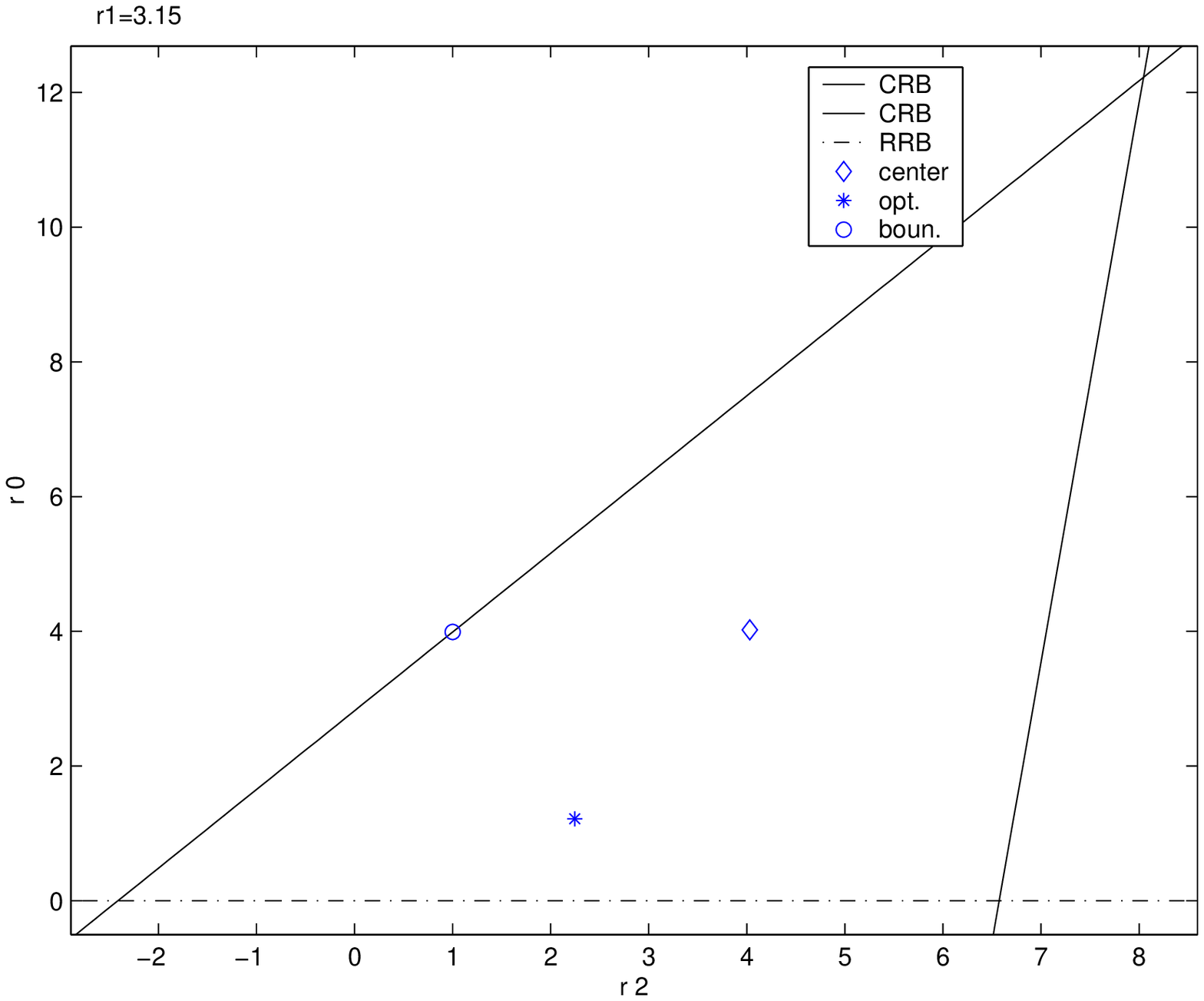}
\caption{Plot of the region when $r_1^*=3.15$}
\label{fig:region_r1_315}
\end{minipage}
\end{center}
\end{figure}

In order to determine complex root boundaries (CRB), we should
first decide, over which interval we should sweep fixed $r_1$. If
we acquire for (\ref{eq:gw}), considering the values for $C$, $N$,
$W_0$, $R_0$, we obtain Figure \ref{fig:r1_interval}. The
interval, in which maximum number of $\omega_{gi}$ is produced,
can be better observed when we look in the interval,
$\omega\in[0,120]$ as in Figure \ref{fig:r1_interval_zoom}. As we
easily observed from Figure \ref{fig:r1_interval_zoom} that it is
enough to sweep $r_1$ between $[-2,8]$ in our problem. Therefore
we obtained the boundary lines for stability on the $(r_0,r2)$
plane for each fixed $r_1$. These boundary lines yield a polygon
in which we have stability. After sweeping $r_1$ and combining all
polygons, we obtain the stability space for controller parameters.

\vspace{.2in}
\centerline {
\includegraphics[width=2in]{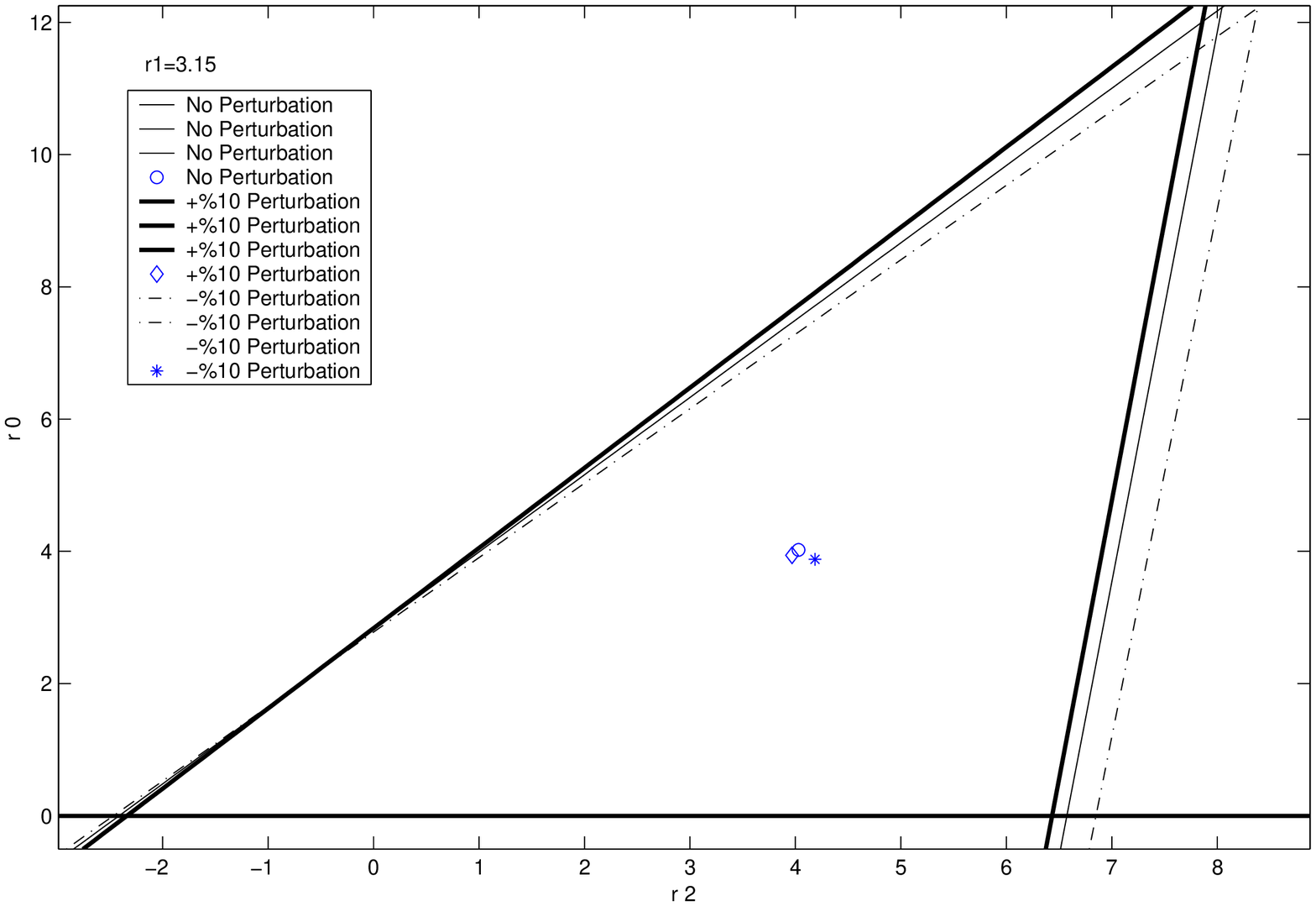}
}
\vspace{.2in}
The region shown in Figure~\ref{fig:region_r1_315} is robust to
small variations (about 10\%) in the original problem data, $N$, $T_p$, $C_n$,
as shown above.

In order to find the optimal PID parameters, we define the cost
function, $\Psi$, with weight functions,
\bea
\nonumber W_1(s)&=&\frac{1+0.01s}{0.01+s}, \\
\nonumber W_2(s)&=&s+1.
\eea

In Figure \ref{fig:r1_vs_cost}, for fixed $r_1\in[-2,8]$, the
minimum value of cost function in $(r_0,r_2)$ plane is given. The
minimum value is achieved at $r_{1,opt}=3.15$. Figure
\ref{fig:region_r1_315} shows the controller parameter space in
$(r_0,r_2)$ plane when $r_1=3.15$. The optimal point is found as
$r_{1,opt}=3.15$, $r_{2,opt}=2.2460$ and $r_{0,opt}=1.2189$. These
normalized values correspond to the $K_{P.opt}=3.845\;10^{-3}$,
$K_{D,opt}=2.091\;10^{-3}$ and $K_{I,opt}=1.48\;10^{-3}$. The
location of optimal, center and one of the boundary points can be
seen in Figure \ref{fig:region_r1_315}.

\subsection{Performance and Robustness of the Feedback System}
\begin{figure}[ht]
\begin{center}
\begin{minipage}[b]{0.5\textwidth}
\centering
\includegraphics[width=3.1in,height=2.5in]{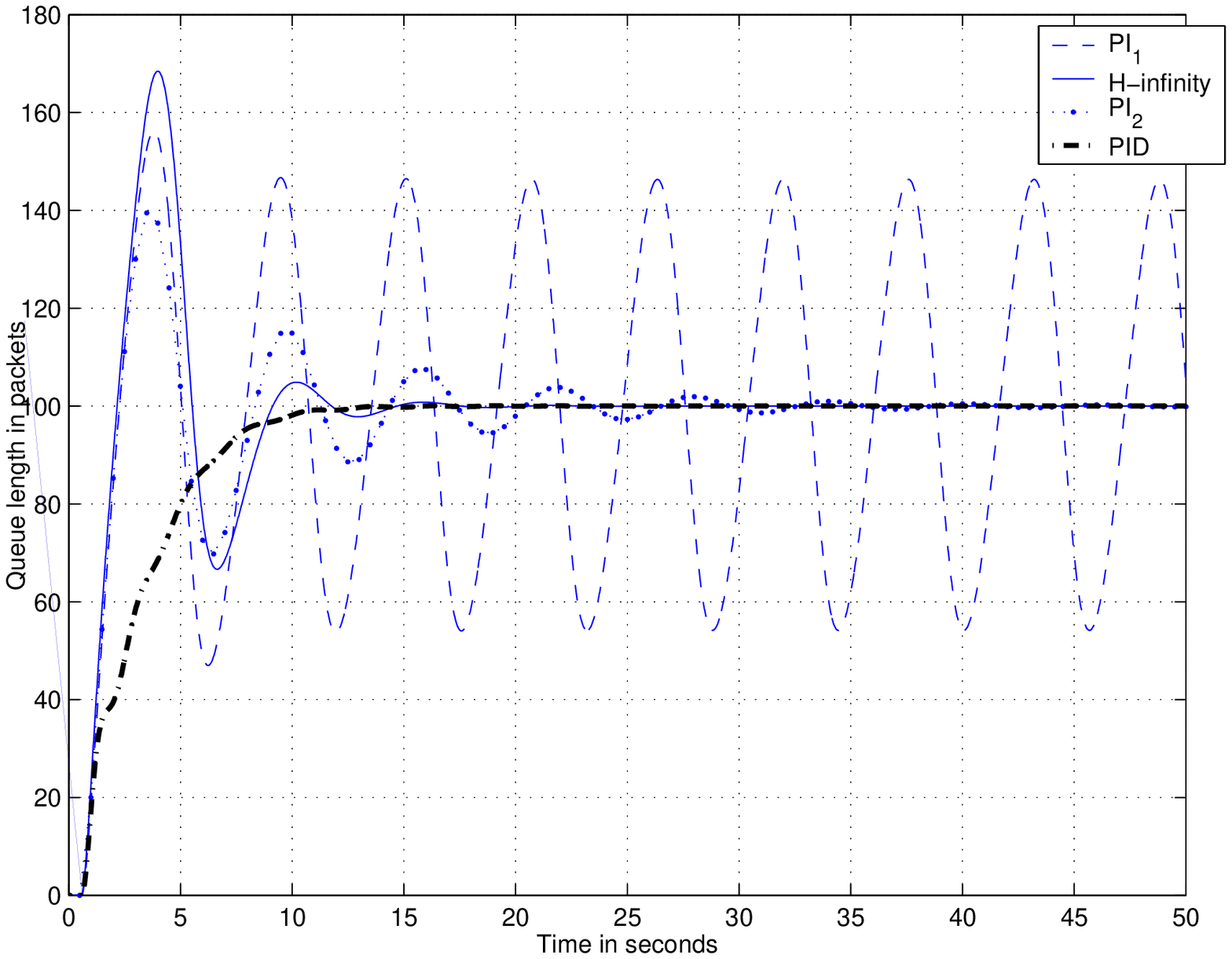}
\caption{Performance comparison of PID, $\Hi$, $PI_1$ and $PI_2$
controllers} \label{fig:stab_r1opt}
\end{minipage}%
\begin{minipage}[b]{0.5\textwidth}
\centering
\includegraphics[width=3in,height=2.45in]{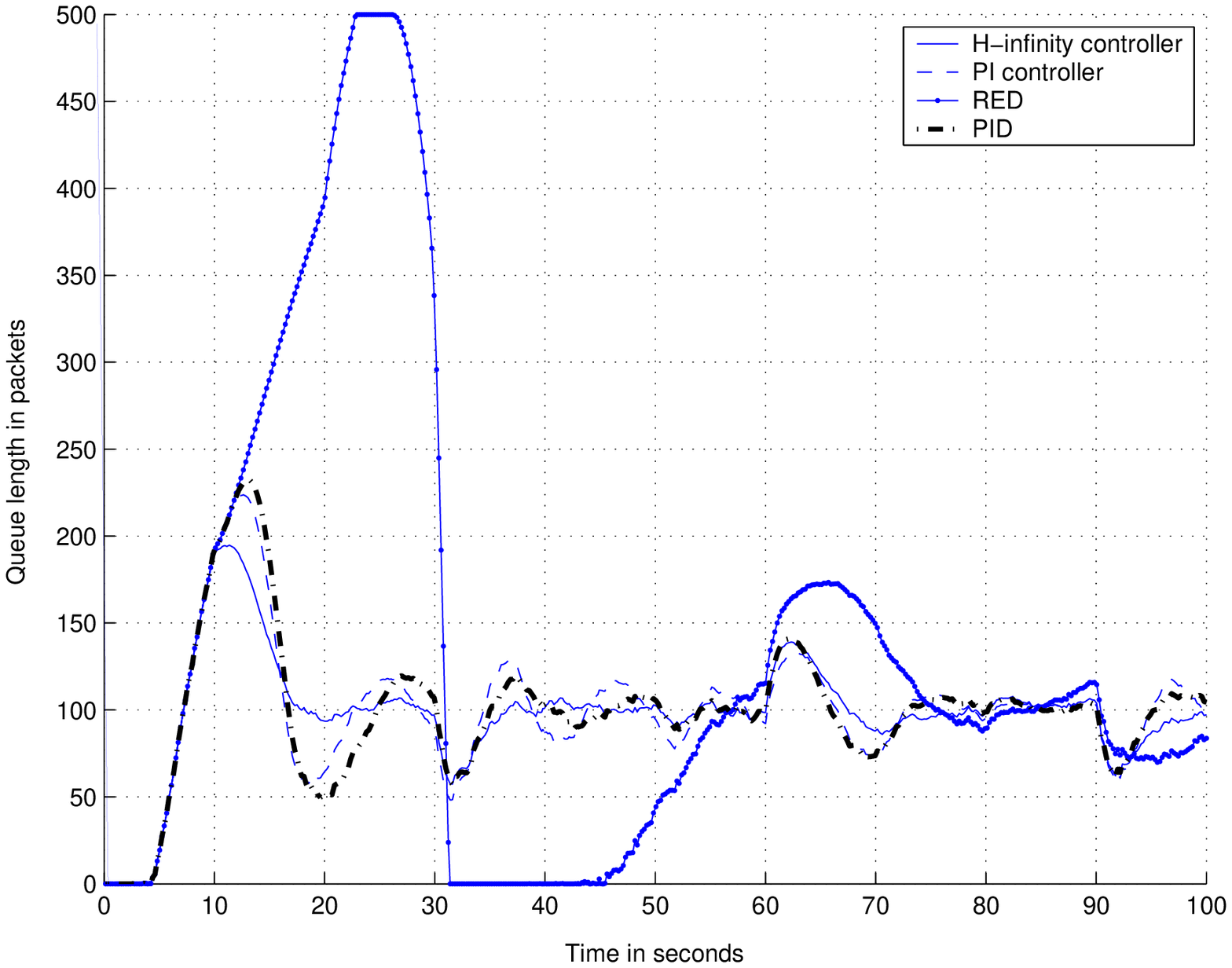}
\caption{Robustness comparison of PID, $\Hi$, $PI_1$ and $PI_2$
controllers} \label{fig:robust_r1opt}
\end{minipage}
\end{center}
\end{figure}

Once the optimal point is found, we need to simulate the feedback system
to check the performance of our controller. In
\cite{QO03}, the performance of $\Hi$ and PI controllers are
compared. Using the simulation parameters of \cite{QO03} (given
above), we obtained Figure \ref{fig:stab_r1opt}, from which the
comparison between our PID, $\Hi$ and $PI_1$ and $PI_2$
controllers can be made.

Figure \ref{fig:stab_r1opt} reveals that PID controller responds
better than other controllers. Although rise time is longer,
settling time of PID is shorter than the other ones. Also note
that there is no overshoot for the proposed PID controller.

\begin{figure}[ht]
\begin{center}
\includegraphics[width=4.1in,height=2.5in]{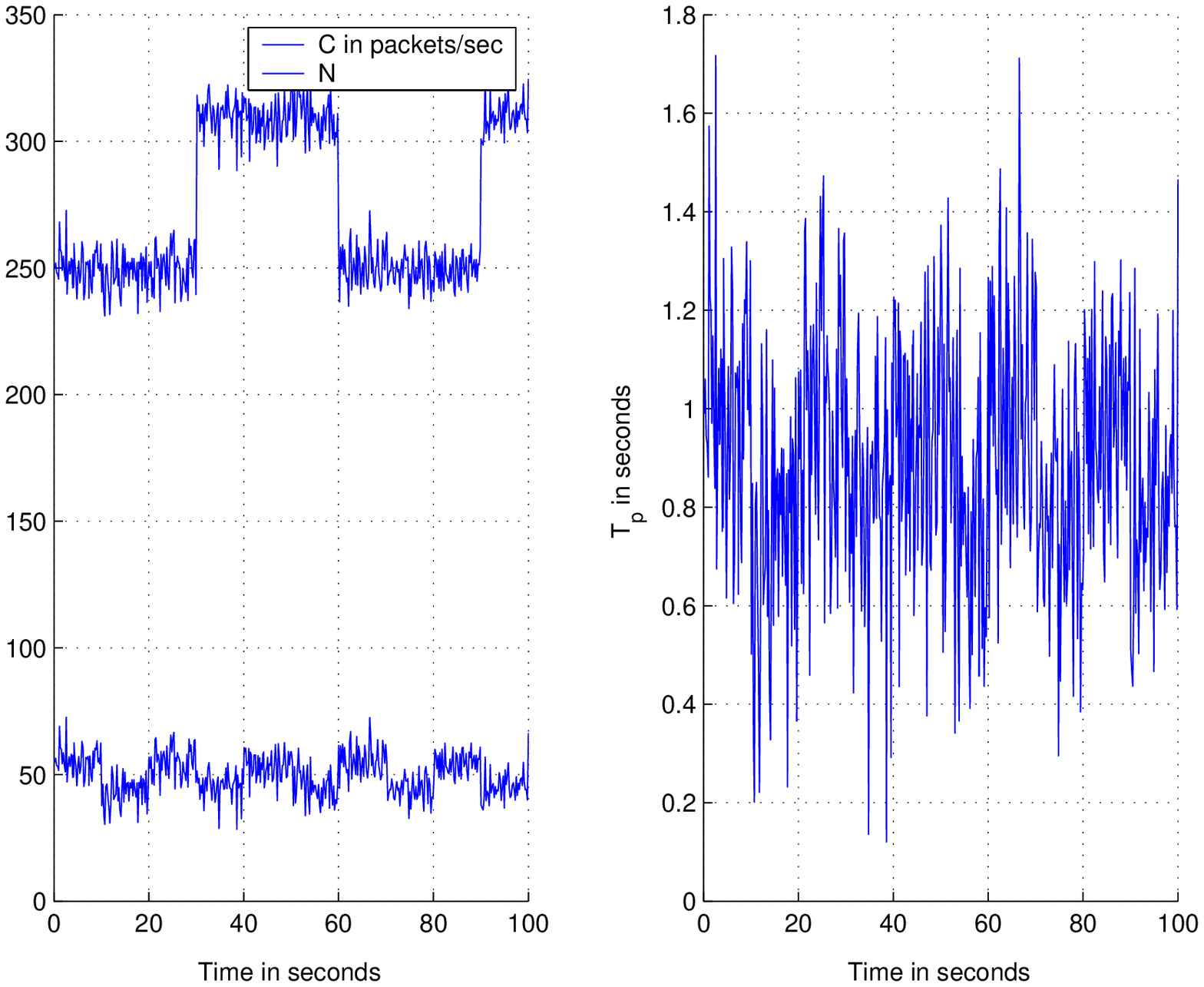}
\caption{Values of $C$, $N$ and $T_p$ corresponding to Figure
\ref{fig:robust_r1opt}} \label{fig:traffic_r1opt}
\end{center}
\end{figure}

For variation in network parameters as shown in Figure
\ref{fig:traffic_r1opt}, robust performance of our global optimum
point is obtained in Figure \ref{fig:robust_r1opt}. We observe
that the PID controller which we design using the method
introduced in \cite{HA03} has similar robust performances with
other proposed controllers of \cite{HMTG02,QO03}.
\subsection{Remarks}
\begin{figure}[h!]
\begin{center}
\begin{minipage}[b]{0.5\textwidth}
\centering
\includegraphics[width=3.1in,height=2.5in]{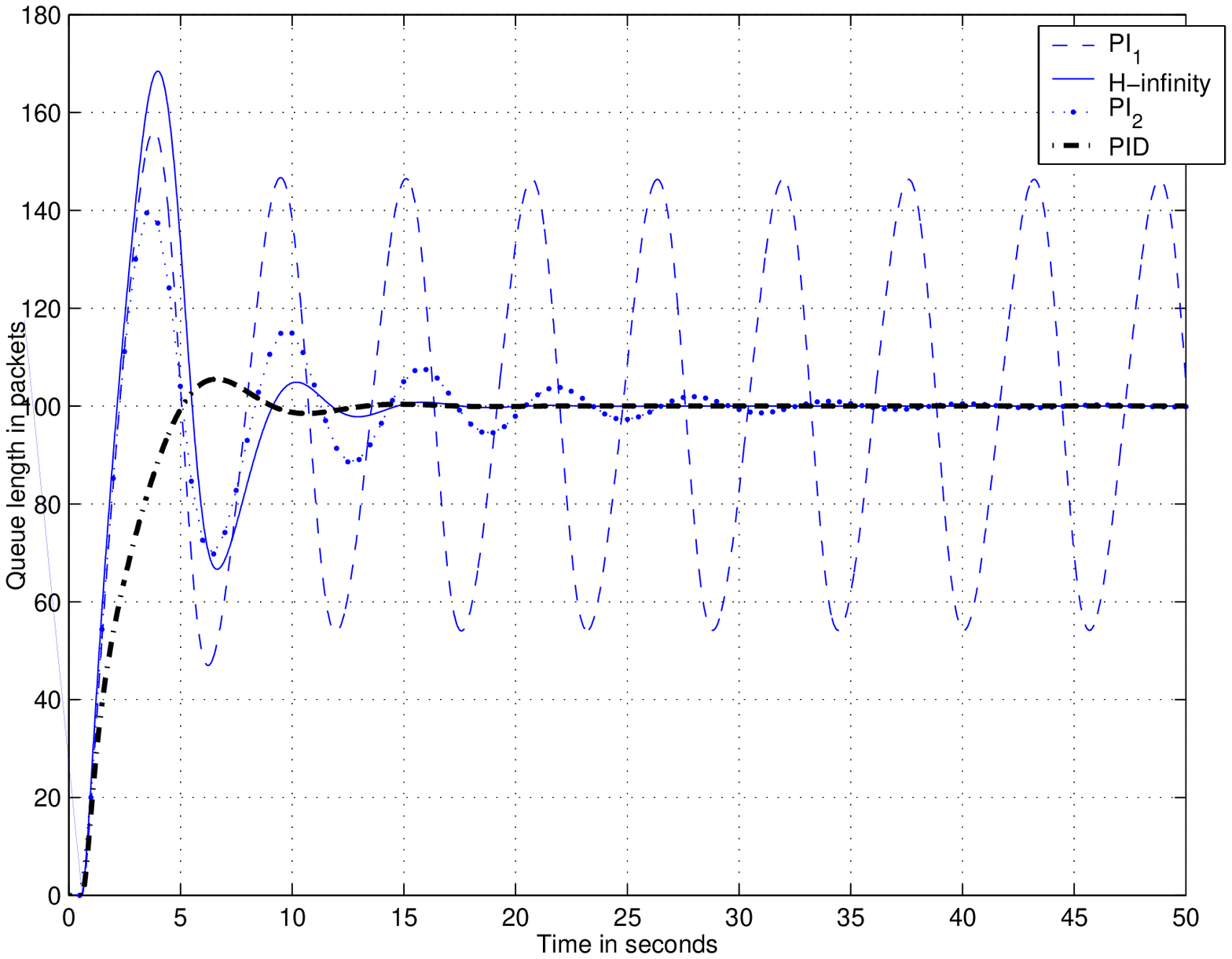}
\caption{Performance comparison of when $r_1=1$, $r_2=0.7016$ and
$r_0=0.839$} \label{fig:stab_r1_1}
\end{minipage}%
\begin{minipage}[b]{0.5\textwidth}
\centering
\includegraphics[width=3in,height=2.45in]{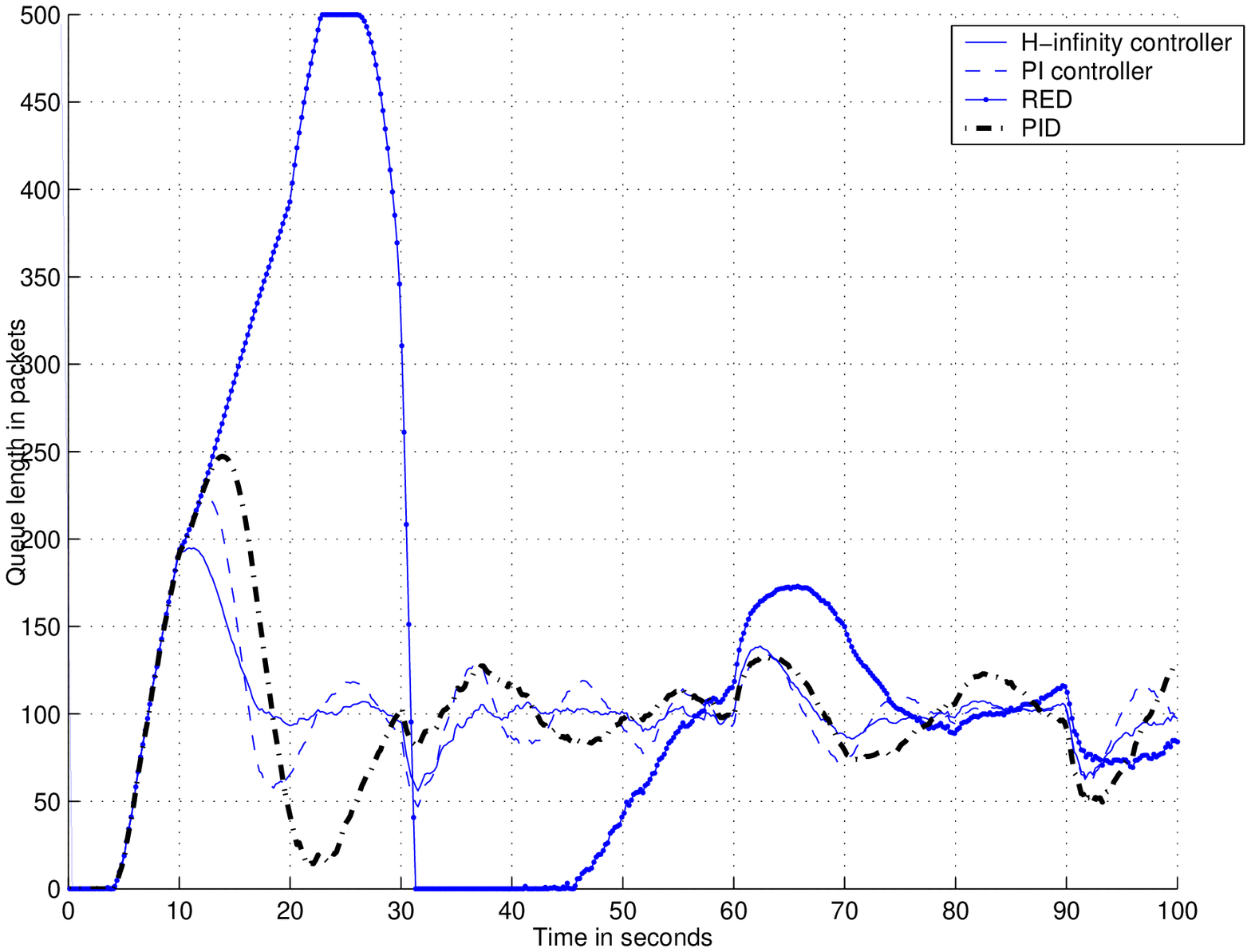}
\caption{Robustness comparison of when $r_1~=~1$, $r_2=0.7016$ and
$r_0=0.839$} \label{fig:robust_r1_1}
\end{minipage}
\end{center}
\end{figure}

\textbf{1)} Since PID controller design is based on linearization
of nonlinear plant, we may encounter different points in the
stable space which give us better performance and robustness.
For example, in our simulations, the results of a PID controller with
parameters $r_1=1$, $r_2=0.7016$ and $r_0=0.839$
($K_P=1.221\;10^{-3}$, $K_D=2.054\;10^{-4}$ and
$K_I=1.024\;10^{-3}$) are given in Figure \ref{fig:stab_r1_1} and
\ref{fig:robust_r1_1}. It can be seen that the controller has a
better settling and rise time with an overshoot. However, the
robust performance of optimal point is better than the point
($r_1=1$, $r_2=0.7016$ and $r_0=0.839$).

\textbf{2)} For confirmation we performed several other
simulations for the following points in the parameter space: (i) center of the
stability polygon, (ii) the boundary of the stability polygon
(shown with diamond and circle symbol in
Figure~\ref{fig:region_r1_315}), which we think intuitively that,
they yield stable and unstable responses, respectively.

\begin{figure}[h!]
\begin{center}
\begin{minipage}[b]{0.5\textwidth}
\centering
\includegraphics[width=3.1in,height=2.5in]{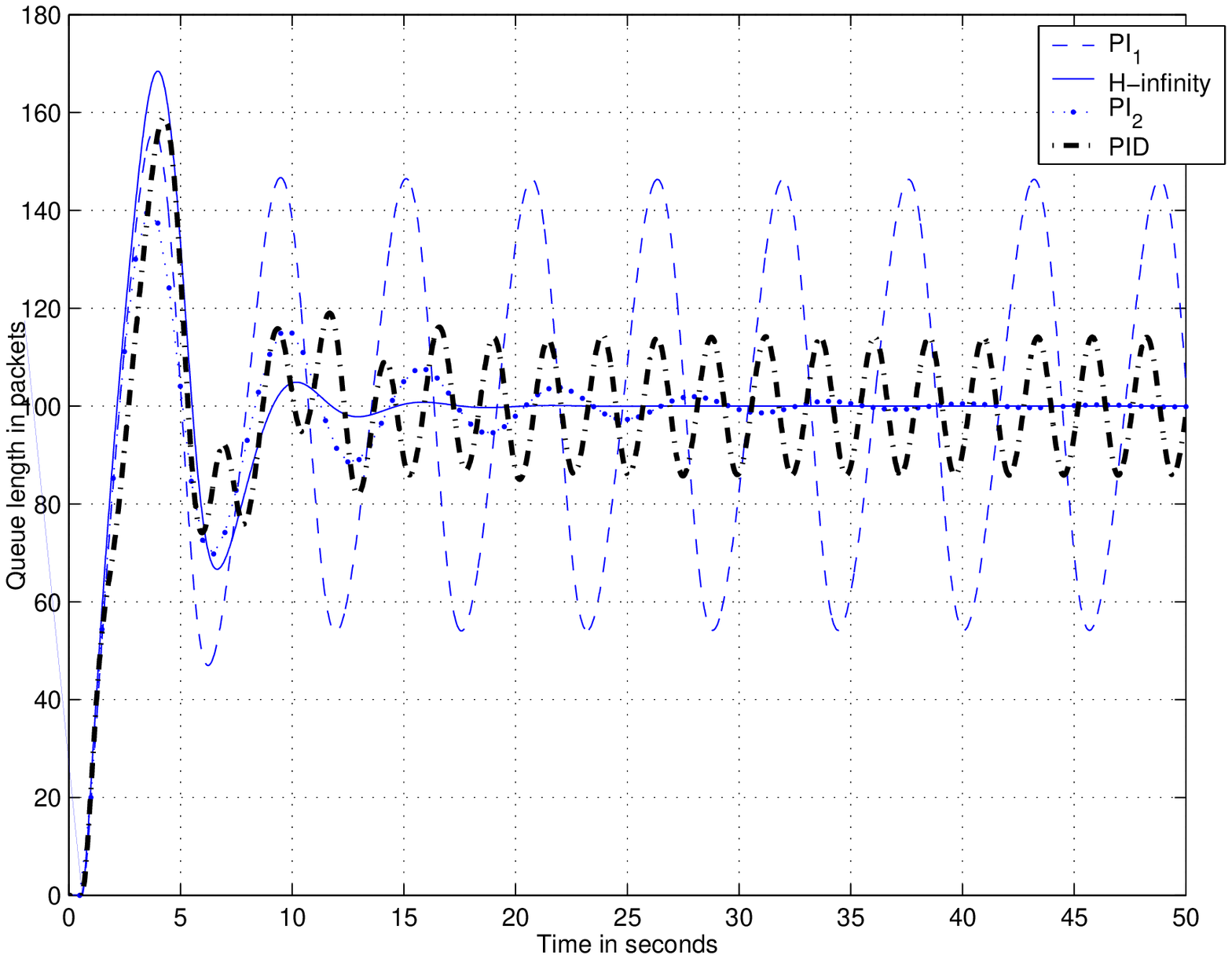}
\caption{Performance comparison of central point}
\label{fig:stab_r1_315_cen}
\end{minipage}%
\begin{minipage}[b]{0.5\textwidth}
\centering
\includegraphics[width=3in,height=2.45in]{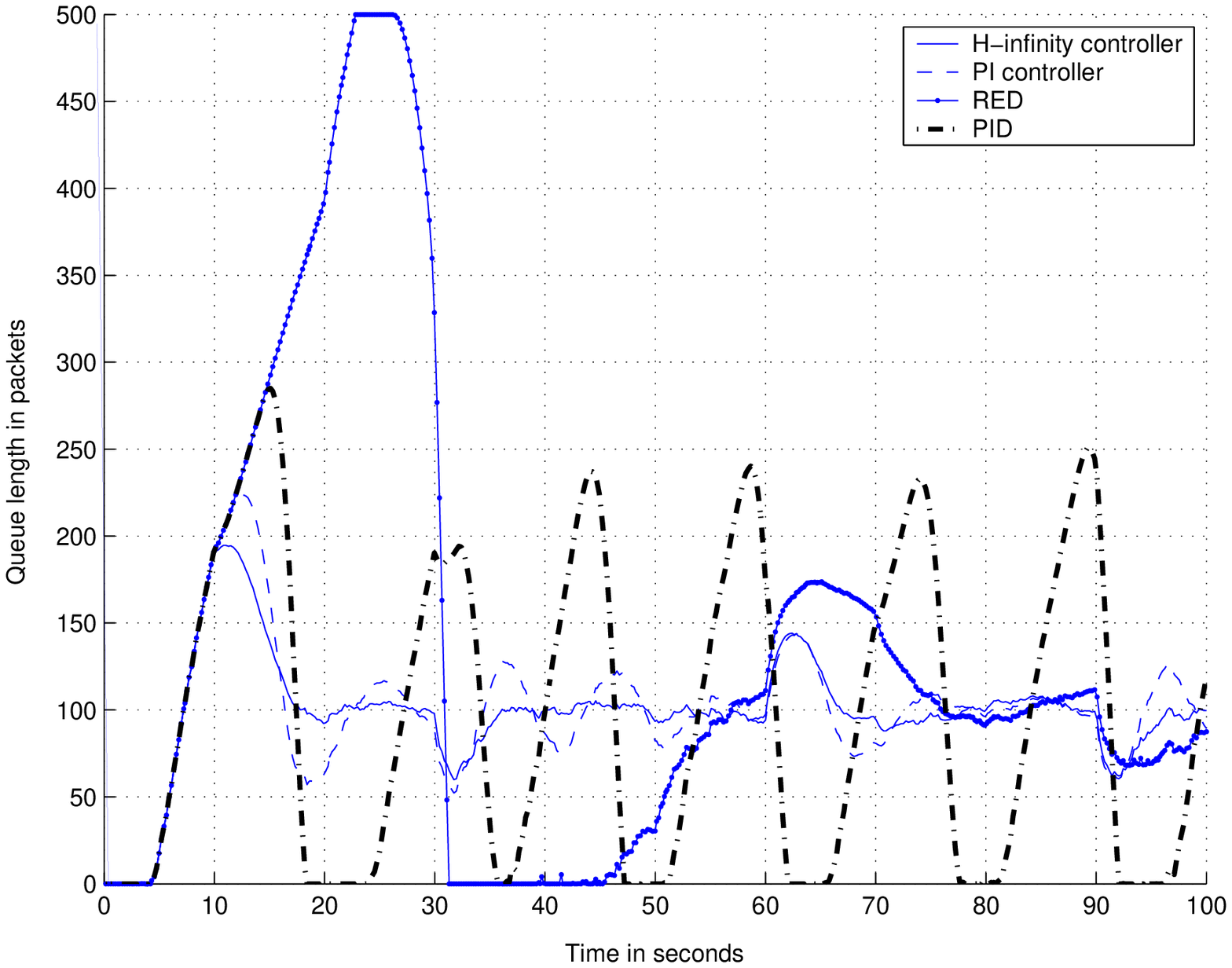}
\caption{Robustness comparison of central point}
\label{fig:robust_r1_315_cen}
\end{minipage}
\begin{minipage}[b]{0.5\textwidth}
\centering
\includegraphics[width=3in,height=2.35in]{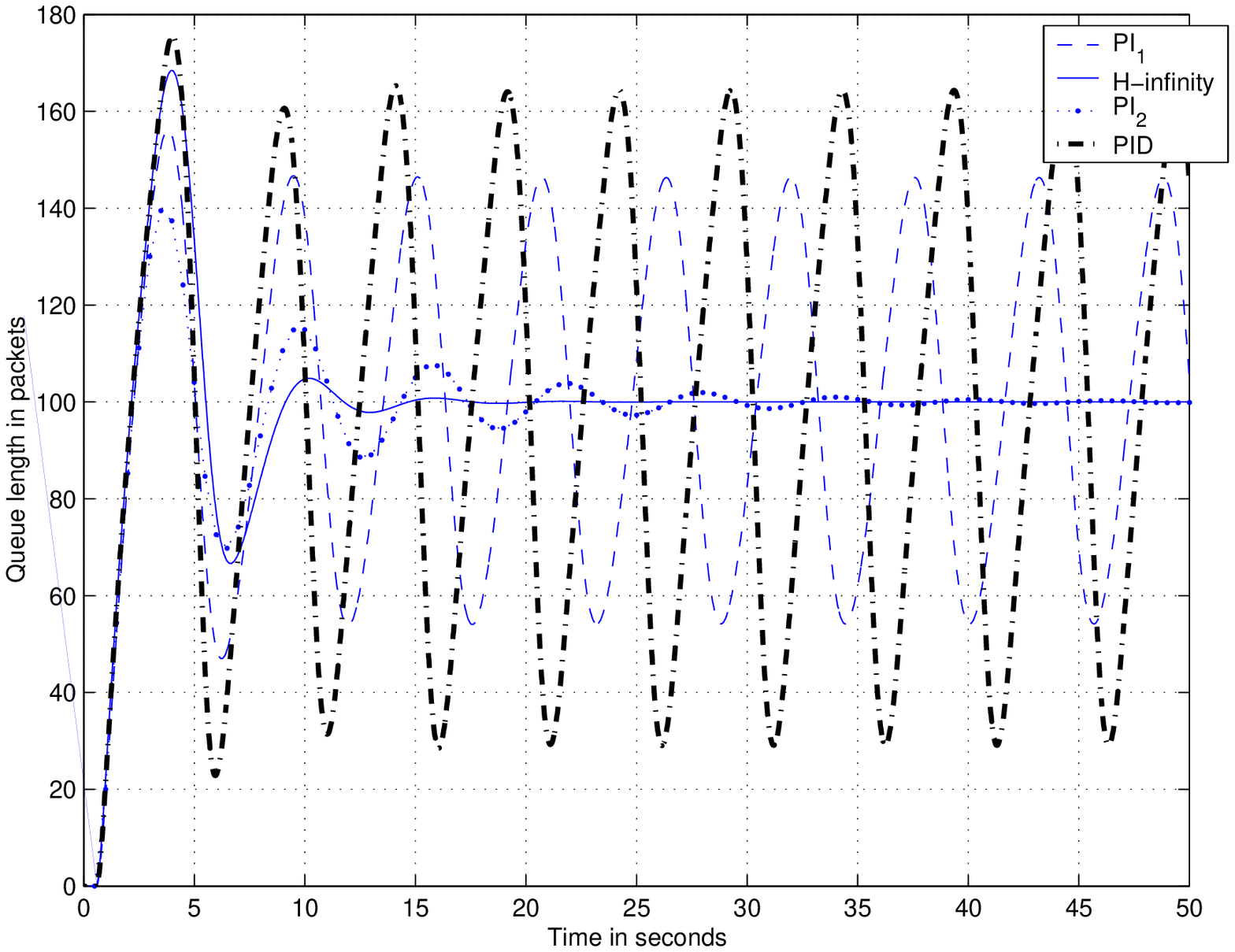}
\caption{Performance comparison of boundary point}
\label{fig:stab_r1_315_boun}
\end{minipage}%
\begin{minipage}[b]{0.5\textwidth}
\centering
\includegraphics[width=3in,height=2.45in]{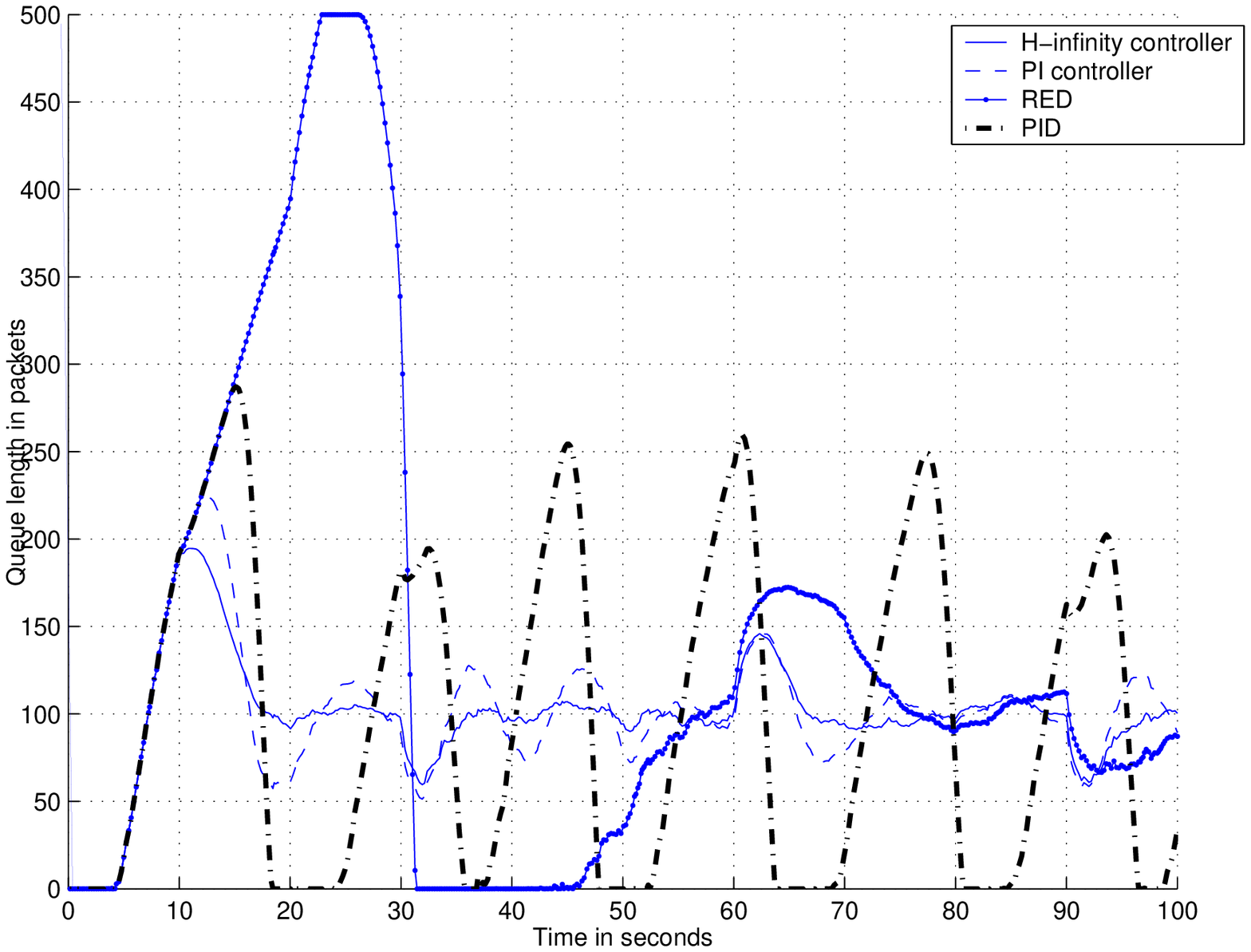}
\caption{Robustness comparison of boundary point}
\label{fig:robust_r1_315_boun}
\end{minipage}
\end{center}
\end{figure}

Figure \ref{fig:stab_r1_315_cen} and \ref{fig:stab_r1_315_boun}
show the responses of the center and  the boundary point
respectively for the r1=3.15 polygon. We can see that stability is
violated as we move to the boundary, which is naturally expected.
This violation can also be observed from the robust performance of
the boundary in \ref{fig:robust_r1_315_boun}. The robust
performance difference is very significant when we compare Figure
\ref{fig:robust_r1opt} and \ref{fig:robust_r1_315_boun}. For the
boundary, the robust response in queue length deviates in
$[0,250]$, unlikely for the optimal point, this deviation is in
$[50,140]$.

\section{Concluding Remarks} \label{sec:remarks}

We proposed a PID controller for robust AQM control scheme
supporting TCP flows. Tuning algorithm for this PID controller is
based on \cite{HA03,OS03} and a numerical search algorithm for
minimization of mixed sensitivity cost function. We compared our
controller performance and robustness with other controllers
studied in \cite{HMTG02,QO03}. For the application on AQM
supporting TCP flows, we obtained relatively good performances
compared to RED, $PI_1$ and $PI_2$ controllers by achieving fast
transients and low oscillatory behavior.

\end{document}